\documentstyle[twoside,fleqn,espcrc2,epsf]{article}
\newcommand{\AmS}{{\protect\the\textfont2
  A\kern-.1667em\lower.5ex\hbox{M}\kern-.125emS}}

\def\lQ{\Lambda_{QCD}}
\newcommand{\nn}{\nonumber}
\newcommand{\be}{\begin{equation}}
\newcommand{\ee}{\end{equation}}
\newcommand{\bea}{\begin{eqnarray}}
\newcommand{\eea}{\end{eqnarray}}
\def\al{\alpha}
\def\als{\alpha_{s}}
\def\siml{{\
    \lower-1.2pt\vbox{\hbox{\rlap{$<$}\lower6pt\vbox{\hbox{$\sim$}}}}\ }} 

\title{Heavy Quarkonium and Nonrelativistic Effective Field Theories}

\author{Antonio Pineda\address{Theory Division CERN, 1211 Geneva 23,
    Switzerland}%
        \thanks{Marie Curie Fellow, contract No. ERBFMBICT983405.}
        }

\begin{document}

\begin{abstract}
We study some general aspects of the formalism of potential NRQCD (pNRQCD), an
effective field theory that
deals with ultrasoft degrees of freedom in Heavy Quarkonium systems. Specific 
attention is paid to its effective Lagrangian that it is displayed at the present level of accuracy. 
\end{abstract}

\maketitle

\section{Introduction}

The last years have witnessed important progress in the theoretical
understanding of heavy-quark--antiquark systems near threshold through the use
of effective field
theories \cite{Lepage,Manohar,pNRQCD,BS98,BV1,short,long} (see also \cite{Labelle,pos,CMY} for
related work in QED). The key point relies on the fact that, since the quark
velocity $v$ is a small quantity, $v \ll 1$, a hierarchy of widely separated
scales: $m\gg mv \gg mv^2 ...$ is produced in these systems, where $m$ is the
heavy quark mass (hard scale), $mv$ the soft scale and $mv^2$ the ultrasoft scale. One can take
advantage of this hierarchy by integrating out the scales above the energies we want our
theory to describe, i.e. above the ultrasoft ones. 

After integrating out the hard scale Non-relativistic QCD (NRQCD) is obtained
\cite{Lepage}. The Lagrangian of NRQCD can be organised in powers of $1/m$.  
After integrating out the soft scale in NRQCD, 
potential NRQCD (pNRQCD) is obtained \cite{pNRQCD}. The Lagrangian of pNRQCD is organised 
in powers of $1/m$ and the relative coordinate ${\bf r}$ (multipole expansion). The matching coefficients in this latter case are
non-analytical functions of ${\bf r}$.

The integration of degrees of freedom is done in practice through a 
matching procedure (see \cite{Manohar,Match,pos,long} for details). The matching 
from QCD to NRQCD can always be done perturbatively since, by definition of heavy quark, $m\gg \lQ$ \cite{Manohar,Match}. 
The matching from NRQCD to pNRQCD can only be carried out perturbatively when
$mv \gg \lQ$. We will assume this to be so throughout this work. Therefore,
the matching coefficients in both NRQCD and pNRQCD can be computed order by
order in $\als$. The non-analytical behaviour in $1/m$ appears through logs in
the matching coefficients of the NRQCD Lagrangian:
$$
c \sim A\als(\ln{m \over \mu_h}+B),
$$
where $\mu_h$ denotes the matching scale between QCD and NRQCD. A typical
matching coefficient for pNRQCD could have the following structure:
\be
 D \sim V({\bf r},{\bf p},{\bf s}_1,{\bf
   s}_2)\left(A^{\prime}\ln{m\,r}+B^{\prime}\ln{\mu\,r}+C\right)
,
\label{coefpNRQCD}
\ee
where $\mu$ denotes the matching scale between NRQCD and pNRQCD and $V$
denotes a non-analytic function of ${\bf r}$.

\section{Theoretical Framework}
\label{sectheory}

\subsection{pNRQCD: the degrees of freedom}
Integrating out the soft scale, $mv$, in  NRQCD produces pNRQCD \cite{pNRQCD}. 
The relevant degrees of freedom of pNRQCD will depend in general on the nonperturbative features of NRQCD. 
In this work we will assume that there exists a matching scale $\mu$ such that $mv\gg\mu\gg mv^2,\, \lQ$, where 
a perturbative picture still  holds. 

Strictly speaking pNRQCD has two ultraviolet (UV) cut-offs $\Lambda_1$ and $\Lambda_2$. 
The former fulfils the relation  $ mv^2,\lQ$  $\ll \Lambda_1 \ll$  $mv$ and is the cut-off of the
energy of the quarks and of the energy and the momentum of the gluons (it corresponds to the $\mu$ above), 
whereas the latter fulfils $mv \ll \Lambda_2 \ll m$ and is the cut-off of the relative
momentum of the quark--antiquark system, ${\bf p}$. In principle, we have some
freedom in choosing the relative importance between $\Lambda_1$ and $\Lambda_2$. 
Our choice is $\Lambda_2^2/m \ll  \Lambda_1$, which guarantees that the UV behaviour 
of the quark propagator in pNRQCD is that of the static one. 

If we denote any scale below $\Lambda_1$, i.e. $mv^2,\lQ$, ..., with  $\Lambda_{mp}$, 
we are in a position to enumerate the effective degrees of freedom of pNRQCD.
These are: $Q$-$\bar Q$ states with energy of $O(\Lambda_{mp})$ and relative momentum not larger 
than the soft scale, and gluons with energy and momentum of $O(\Lambda_{mp})$. 
Let us define the centre-of-mass coordinate of the $Q$-$\bar Q$ system 
${\bf R} \equiv ({\bf x}_1+{\bf x}_2)/2$ and the relative coordinate ${\bf r} \equiv {\bf x}_1-{\bf x}_2$. 
A $Q$-$\bar Q$ state can be decomposed into a singlet state $S({\bf R},{\bf r},t)$ and an octet 
state $O({\bf R},{\bf r},t)$, in relation to colour gauge transformation 
with respect to the centre-of-mass coordinate. We notice that in QED only the analogous to the singlet appears. 
The gauge fields are evaluated in ${\bf R}$ and $t$, i.e.  $A_\mu = A_\mu({\bf R},t)$: 
they do not depend on ${\bf r}$. This is due to the fact that, since the typical size  ${\bf r}$ 
is the inverse of the soft scale, gluon fields are multipole expanded with respect to this variable.

\subsection{Power counting}
In order to discuss the general structure of the pNRQCD Lagrangian, let us consider  in more detail 
the different scales involved in the problem 
$$
m\,, \quad p\,, \quad {1 \over r} \, ,\quad \Lambda_{mp}\,.
$$
The variables $r$ and $m$ will explicitly appear in the pNRQCD Lagrangian, since they 
correspond to scales that have been integrated out. Although $p$ and $1 / r$ are of the same size 
in the physical system, we will keep them as independent. This will facilitate the counting rules 
used to build the most general Lagrangian. 

With the above objects, the following small dimensionless quantities will
appear: 
\begin{equation}
{p \over m},\quad {1 \over r m}, \quad \Lambda_{mp}r \quad \ll 1.
\label{uneq1}
\end{equation} 
Note that $1/p r$ or $\Lambda_{mp} /p$ are not allowed since $p$ has to appear in an analytic way 
in the Lagrangian (this scale has not been integrated out). 
The last inequality in Eq. (\ref{uneq1}) tells us that $r$ can be considered
to be small 
with respect to the remaining dynamical lengths in the system. As a consequence the 
gluon fields can be systematically expanded in $r$ (multipole expansion).

Therefore, the pNRQCD Lagrangian can be written as an expansion in $1/m$ 
(from the first two inequalities of Eq. (\ref{uneq1})), 
and as an expansion in $r$ (the so-called multipole expansion, corresponding to the 
third inequality of Eq. (\ref{uneq1})).

\subsection{pNRQCD: the Lagrangian}

The pNRQCD Lagrangian would look as follows:
\begin{eqnarray}                         
&&{\cal L}_{\rm pNRQCD} =
{\rm Tr} \,\Biggl\{ {\rm S}^\dagger \left( i\partial_0 
- {{\bf p}^2\over m} +{{\bf p}^4\over 4m^3}
\right.
\nonumber
\\
&&
\left. \qquad
- V^{(0)}_s(r) - {V_s^{(1)} \over m}- {V_s^{(2)} \over m^2}+ \dots  \right) {\rm S}
\nonumber
\\
&& \qquad 
\nonumber 
 + {\rm O}^\dagger \left( iD_0 
- {{\bf p}^2\over m}
- V^{(0)}_o(r) 
+\dots  \right) {\rm O} \Biggr\}
\nonumber\\
\nonumber
& &\qquad + g V_A ( r) {\rm Tr} \left\{  {\rm O}^\dagger {\bf r} \cdot {\bf E} \,{\rm S}
+ {\rm S}^\dagger {\bf r} \cdot {\bf E} \,{\rm O} \right\} 
\\
\nonumber
& &\qquad
+ g {V_B (r) \over 2} {\rm Tr} \left\{  {\rm O}^\dagger {\bf r} \cdot {\bf E} \, {\rm O} 
+ {\rm O}^\dagger {\rm O} {\bf r} \cdot {\bf E}  \right\}  
\\
& &\qquad- {1\over 4} F_{\mu \nu}^{a} F^{\mu \nu \, a}\,,
\label{pnrqcdph}
\end{eqnarray}
where the dots indicate higher-order potentials in the $1/m$ expansion, we
have neglected center-of-mass variables and we have just kept $O(r)$ terms in
the multipole expansion.  
Let us now display the structure of the potentials up to $O(1/m^2)$. 

\noindent
{\bf Order $1/m^0$}. By dimensions $V^{(0)}_s(r)$ can only have the 
following structure:
\begin{equation}
V^{(0)}_s(r) \equiv  - C_f {\alpha_{V^{(0)}_s}(r) \over r}, \quad C_f={N_c^2-1
  \over N_c},  
\label{defpot0}
\end{equation}
{\bf Order $1/m$}. By dimensions plus time reversal $V^{(1)}_s(r)$ can only 
have the 
following structure:
\be
{V^{(1)}_s \over m} \equiv -{C_fC_A D^{(1)}_s \over 2mr^2}, \quad C_A=N_c .
\ee
{\bf Order $1/m^2$}. The following matching potentials appear to the accuracy
to which the matching has been performed at present 
\bea
&&{V^{(2)}_s \over m^2} = 
- { C_f D^{(2)}_{1,s} \over 2 m^2} \left\{ {1 \over r},{\bf p}^2 \right\}
+ { C_f D^{(2)}_{2,s} \over 2 m^2}{1 \over r^3}{\bf L}^2
\nn
\\
&&
+ {\pi C_f D^{(2)}_{d,s} \over m^2}\delta^{(3)}({\bf r})
+ {4\pi C_f D^{(2)}_{S^2,s} \over 3m^2}{\bf S}^2 \delta^{(3)}({\bf r})
\nn
\\
&&
+ { 3 C_f D^{(2)}_{LS,s} \over 2 m^2}{1 \over r^3}{\bf L} \cdot {\bf S}
+ { C_f D^{(2)}_{S_{12},s} \over 4 m^2}{1 \over r^3}S_{12}(\hat {\bf r})
.
\label{V2}
\eea
Note that ${\bf p}$ appears analytically in the potentials. The power on $n$
(${\bf p}^n$) to which ${\bf p}$ appears in the potential is constrained by the power in $1/m$. 
Some of the operators above become too singular once the dependence on $\ln r$
is taken into account in the matching coefficients. For instance, 
one could have $\delta({\bf r})\ln{m r}$ which is ill defined. In fact, this
operator has to be understood as the sum of two operators, well defined each:
$\delta({\bf r})\ln{m \over \mu_h}$ and $reg{1 \over r^3} +\ln{\mu_h}$. The
latter is basically the Fourier transform of $\ln{\mu_h \over k}$ (see
\cite{pos} and references therein). 

The matching potentials for the octet sector could be defined in a similar way
by changing the overall factor $C_f$ by $C_f-C_A/2$ and the subscript $s$ by
$o$ in each of the matching coefficients ($\al_{V_o}$, $D^{(n)}_o$).

The above representation of the potential can be related with others found in the 
literature \cite{Gupta} by the use of the equations of motion.

In order to have the proper free-field normalisation in the colour space we define $
{\rm S} \equiv { 1\!\!{\rm l}_c \over \sqrt{N_c}} S $ and ${\rm O} \equiv  { T^a \over \sqrt{T_F}}O^a, $
where $T_F=1/2$, and $S$ and $O$ are $1/2\times 1/2$ tensors in spin space.

In principle, terms with higher time derivatives acting on the fields 
$S$ or $O$ could be considered in the pNRQCD Lagrangian. These terms are 
redundant in the sense that one could make them disappear from the Lagrangian and still correctly predict 
all physical observables by re-shuffling the values of the matching coefficients. We have chosen the minimal form of the pNRQCD Lagrangian, where higher time derivatives are absent. In fact, one can always get rid of these terms by systematically using 
local field redefinitions without changing the physical observables 
(spectrum) of the theory. 

\subsection{pNRQCD matching coefficients}

In order to obtain the different matching coefficients in pNRQCD:
${\alpha}_{V_s}$, $D^{(1)}$, $D^{(2)}$ ... one has to perform the
matching between NRQCD and pNRQCD. A detailed description of the procedure can
be found in \cite{pNRQCD,pos,long,BPSVpre}. Here we just state the main
results and briefly discuss how they are obtained. 
The matching coefficients read as follows
$$
{\alpha}_{V_s}(r, \mu)=\alpha_{\rm s}(r)
\left\{1+\left(a_1+ 2 {\gamma_E \beta_0}\right) {\alpha_{\rm s}(r) \over 4\pi}\right.
$$
$$
+\left[\gamma_E\left(a_1\beta_0+ {\beta_1 \over 2}\right)+\left( {\pi^2 \over 12}+\gamma_E^2\right) 
{\beta_0^2}+{a_2 \over 4}\right] {\alpha_{\rm s}^2(r) \over 4\,\pi^2} 
\nonumber
$$
\be
\left.
+ {C_A^3 \over 12}{\alpha_{\rm s}^3(r) \over \pi} \ln{ r \mu}\right\};
\label{newpot0}                  
\ee
\be
D^{(1)}_s=\alpha_{\rm s}^2(r)\left\{1+{2 \over 3}(4C_f+2C_A){\als \over \pi}
\ln{r\mu} \right\};
\ee
\be
D^{(2)}_{1,s}=\alpha_{\rm s}(r)\left\{1+{4 \over 3}C_A{\als \over \pi}
\ln{r\mu} \right\};
\ee
\be
D^{(2)}_{2,s}=\alpha_{\rm s}(r)
;
\ee
\bea
\nn
&&
D^{(2)}_{d,s}=\alpha_{\rm s}(r)(2+c_D-2c_F^2)
\\
&&
+{1 \over \pi}\left[
d_{vs}+3d_{vv}+
{1 \over C_f}(d_{ss}+3d_{sv})
\right]
\nn
\\
&&
\nn
+
{16 \over 3}{\alpha_{\rm s}^2 \over \pi}({C_A \over 2} -C_f)\ln{r\mu}
\\
\nn
&&
\simeq
\alpha_{\rm s}(r)\left\{1+ 
{\als \over \pi}\left[{2 C_f \over 3} +{17 C_A \over 3}\right]\ln{m r}
\right.
\\
&&
\left.
+ {16 \over 3}{\als \over \pi}({C_A \over 2} -C_f)\ln{r\mu}
\right\}
;
\eea
\bea
&&D^{(2)}_{S^2,s}=\alpha_{\rm s}(r)c_F^2-
{3 \over 2\pi C_f}(d_{sv}+C_f d_{vv})
\nn
\\
&&
\simeq
\als(r)\left(1-{7 C_A \over 4}{\als \over \pi}\ln{m r} \right)
;
\eea
\bea
\nn
&&D^{(2)}_{LS,s}={\alpha_{\rm s}(r) \over 3}(c_S+2c_F)={\alpha_{\rm s}(r)
  \over 3}(4c_F-1)
\\
&&
\simeq
\als(r)\left(1-{2 C_A \over 3}{\als \over \pi}\ln{m r} \right)
;
\eea
\be
D^{(2)}_{S_{12},s}=\alpha_{\rm s}(r) c_F^2
\simeq
\als(r)\left(1-C_A{\als \over \pi}\ln{m r} \right)
.
\ee

In order to obtain $\alpha_{V_s}$ with the accuracy above, it is necessary to
perform the matching between NRQCD and pNRQCD (at $O(1/m^0)$) exactly at the two-loop level
and with the leading-log accuracy at the three-loop level, i.e. to compute the
static potential to this order. The one-loop result was obtained in
Ref. \cite{1loop}, the two-loop one in Ref. \cite{twoloop} and the three-loop
leading-log in \cite{short}. The $\beta_n$ are the coefficients of the beta
function and the values of $a_1$ and $a_2$ can be found in Ref. \cite{twoloop}
(see also \cite{twoloop} for notation). 

For $D_s^{(1)}$ the first
contribution appears at one loop and, to our
knowledge, was first calculated in \cite{Duncan}. The two-loop
leading-log term is obtained in \cite{BPSVpre} (see also \cite{KP}). 

The tree-level result for the different $D_s^{(2)}$ can be obtained from the results of positronium (see
\cite{pos}). The one-loop leading-log dependence on
$\ln{\mu r}$ is obtained in \cite{BPSVpre} (see also \cite{KP}). By taking into account the NRQCD
matching coefficients ($c_D$, $c_F$ ...; see
\cite{Manohar,Match} for its values at one loop) in the vertices when matching to pNRQCD, one obtains the
leading-log dependence on $\ln{mr}$ plus also some finite pieces. In any case,
since the full $O(\als^2)$ correction to the different $D$'s is not yet
available, the results above should only be trusted up to finite pieces at one
loop. It is also worth noting that the spin-dependent matching potentials do
not suffer from ultrasoft divergences at one-loop.

In order to obtain the leading-log dependence on $\mu$ of the matching 
coefficients above it is enough to compute the UV divergences of pNRQCD at
next-to-leading order in the multipole expansion and up to $O(1/m^2)$
\cite{BPSVpre} (see also \cite{KP}).

With the matching coefficients above, the leading-log, $O(m\als^5)$,  
corrections to the heavy quarkonium spectrum can be computed \cite{BPSVpre,KP}.

With little effort a few things can be said with respect to the octet matching
potentials. $\al_{V_o}$ coincides with $\al_{V_s}$ at one loop (although not
necessarily beyond\footnote{We thank Schr\"oder for communication on this
  point.}) and also the leading-log (three-loop) dependence is known
\cite{long}. We are not aware of any calculation of $D^{(1)}_o$ (of which the 
first non-zero contribution also appears at one-loop). The tree-level
contribution to the different $D^{(2)}_o$ can also be read for the positronium
calculation, keeping in mind that, unlike in the singlet case, in this case the annihilation diagram gives a
nonzero contribution (as in QED).

The leading-log dependence of $V_A$ and $V_B$ on $\mu$ is also known
\cite{long}. It reads 
\begin{equation}
V_A(r,\mu) = V_B(r,\mu) = 1 + {8\over 3} C_A {\alpha_{\rm s}\over \pi} \ln{r\mu}.
\label{varun}
\end{equation}

{\bf Acknowledgements}

I thank N. Brambilla, J. Soto and A. Vairo for collaboration on the work
presented here.

\end{document}